# The contrivance of Neptune
## Davor Krajnović

*Celebrating 170th anniversary of the discovery of Neptune, I review the story of the discovery that startled the world. The story is an interplay of scientific triumph and human weakness and an example of how science works in a socio-political context.*

"*Le planète, dont vous avez signalé la position, réellement existe*". This is the opening sentence of an extraordinary letter sent by Johan Gottfried Galle from the Berlin Observatory to Jean Joseph Urabin Le Verrier on September 25, 1846: "The planet whose position you predicted really exists". One can only imagine the emotions of Galle while writing it, or those of Le Verrier reading it three days later in Paris. This sentence announced the most remarkable confirmation of a theoretical prediction in the history of science. It heralded a triumph of Newtonian theory of gravity, astonishing mathematical work, and masterfully executed observations. Neither Galle, nor Le Verrier could have imagined what a storm it would raise.

**The discovery**
The showdown started on the morning of September 23 when Galle, the assistant astronomer to the director Johan Encke, received a letter from Le Verrier. To receive a letter from the eminent French astronomer was surprising, but not fully unexpected for Galle; it was just about a year and a half late. In March 1845, Galle defended a thesis presenting a new reduction of the observations by Ole Rømer in 1706, comprising 88 stars and known planets. As Le Verrier was then trying to calculate the orbit of Mercury, Galle sent him the dissertation knowing the value of such early observations. There was no "thank you" or even an acknowledgement from Le Verrier, perhaps because by that time Le Verrier's focus had shifted to another mystery in the Solar System, the unpredictable motion of Uranus.

Le Verrier's letter started with a delayed thank you, a congratulation on the good work, and a promise to write in more detail about the Mercury issue, but quickly changing the topic to something else: a suggestion to an "indefatigable observer" to look at a very particular place on sky, where a planet could be found. He explained that the location is the result of his work on the irregular motion of Uranus, and provided a very clear location on the sky, as well as a likely size of the planet, which should be resolvable by a good telescope.

This letter was exceptional in many ways. It transmitted a bold, but clear prediction of the location of a new planet, based on Newton's theory of gravity and a complex and novel theory of planetary perturbations, presented some 20 days earlier at a meeting of the Académie des Sciences in Paris. It was a direct solicitation to search for the predicted planet, but it was addressed to an assistant at an observatory some 900 km away in a different country. On top of this, it arrived on the day of the director's 55th birthday!

Today this seems as an amazing opportunity, an insider information, an unmissable tip-off that can secure fame for the recipient and the institution. Astronomy in mid-19th century was, however, very different. State observatories were not research institutes in the present sense, but primarily factories of useful data, from time keeping to charting the skies. The use of the observatory telescopes were under discretion of the director and, as an assistant, Galle had to ask for a permission to observe for his own private research.

The director of the Berlin Observatory, Encke was aware of Le Verrier's theory that a more distant planet perturbs the motion of Uranus, and did not think much of it. But when Galle approached him with the letter, he agreed that it presented a "moral commitment" to Galle to look for the planet. The standard story (e.g. Turner 1911, Grosser 1962, Standage 2000) is that Encke reluctantly gave permissions to Galle to observe that night, but Galle's own account (Galle 1877) is different: while Encke was not much in favour of looking for the planet *before*, once the letter arrived he did not object. He himself didn't want to do it, maybe because it was his birthday, but he gave Galle permission immediately.

Their planning was overheard by another, younger assistant (a student in modern terms), Heinrich Louis d'Arrest, who immediately asked Galle for permission if he could join the observations. And so it was, while the director was celebrating with his family, that Galle and d'Arrest started the search for Le Verrier's planet. As Galle later explained (Galle 1877), the night was clear and they first attempted to look for an object with a clear disc of about 3", but this was not successful. It seemed that they would need to identify all the stars in the area. d'Arrest then suggested to look among the new charts, prepared by Carl Bremiker for the Royal Academy of Sciences in Berlin, to see if one of them covered the area. Galle led the way to Encke's office, where they searched the charts and recognised that a bottom left corner of a chart for the hour XXI covers the region indicated by Le Verrier (see Box 1, "The chart").

Back in the dome, Galle was observing and reading out the positions of stars, while d'Arrest was checking against the chart, until an 8th magnitude star was found that was absent from the chart! One can imagine the silence that followed on that fresh early autumn night, just after midnight, the rechecking of the coordinates, d'Arrest eager to see for himself, Galle double and triple checking the map, the last look in the eyes between the two astronomers, the first to actually see the new planet, just under one minute of arc away from the predicted position. Then they rushed to inform Encke and were all back in the dome to continue observing until the object set. Encke agreed this object had a resolved disc, although somewhat smaller than predicted. The short time left for observing, however, was not enough to detect its motion.

There was nothing for it but to wait until the next night. If it were a planet, and its size was a good indication that it really is, it will not be at the same spot in the sky, and, especially, it would not be a forgotten - and a very bright! - entry in a brand new map from a respectable chart maker. 24 September must have been a very long day for these members of the Berlin Observatory, nervously eyeing the clouds. The night was clear; Galle, d'Arrest and Encke gathered in the dome.

From the start of observations it was evident that the object had moved, that the planet whose position was signalled by Le Verrier, really existed.

**Bringing Uranus under control**
The planet Uranus was discovered by William Herschel in 1781, but he was not the first to see it. There were 17 earlier observationsin which it was considered a star, by J. Flamsteed, T. Mayer, P.C. Lemonnier, and J.J. Lalandel; Lemonnier observed it 11 times over 21 years! These observations were important as they allowed tracing of the planet's motion over a significant part of its orbit. By the start of the 19th century it was clear that there was something a miss with Uranus. Its observed position on the sky was regularly not the same as the predicted one: its behaviour was very peculiar. For example, if one would use only the "modern" observations made after the discovery to determine the orbit of the planet, one could not accommodate the "ancient" observations from before the discovery and vice versa. Furthermore, the discrepancy between the predicted and observed position was increasing towards the turn of the century, just to stabilise and then start decreasing in the 1820s, almost disappear around 1830 and then to suddenly become larger than ever before by 1840s (see Figure 1).

This was a major problem for the usually very precise science of celestial mechanics. Leading astronomers were measuring the deviations and debating their origins. The Astronomer Royal, George Biddle Airy, lead an important observational campaign of Uranus' motion at the Royal Greenwich Observatory (RGO), which later provided crucial data for estimating the position of Neptune. Airy even determined that the distance of Uranus from the Sun (the so-called "radius-vector", which is much more difficult to measure in comparison to the longitudinal displacement), was also changing (Airy 1838). Alexis Bouvard assembled tables of Uranus motion and struggled to bring forward any resolution, even after the influence of Jupiter and Saturn were taken into account.

Such an interesting problem generated several possible solutions. Bouvard himself was of the opinion that something must be wrong with the "ancient" observations, that they are not as precise as the modern ones. This idea was, however, quickly rejected as even the modern observations became discrepant from the predictions soon after the publication of the tables. A similar fate befell a more physical conjecture, that a comet hit Uranus around the time of the discovery, changing its orbit; the continuing changes to the orbit ruled that out, too. Other physical theories involved the existence of a medium through which Uranus moves and slows its motion, or the suggestion that Uranus had a massive moon. Neither was compatible with the data spanning more than a century. There were two final possibilities, either the law of gravity was not the same at those huge distances from the Sun, or there might be another, unseen planet disturbing the orbit of Uranus.

Alternative theories of gravity were not a novelty then, as they are not now, but Newtonian theory of gravity did withstand all tests thrown at it. Le Verrier was never in doubt that Newtonian gravity is correct, and that there could only be one cause for the anomalous motion of Uranus: a new planet.

The first paper dealing with the "Theory of Uranus" was presented by Le Verrier on November 5, 1845 (Le Verrier 1845). It dealt with existing data on the anomalous motion of Uranus, rejecting the claims of Bouvard that the ancient data were wrong, and demonstrating that when the influence of both Saturn and Jupiter is removed, there are significant residuals between the observed and predicted motion (Figure 1). In his second paper, presented on June 1, 1846 to the Académie des Sciences (Le Verrier 1846a), Le Verrier rejected all other theories invoked to explain the motion of Uranus and showed that the new planet could not be interior to the orbit of Uranus. His choice was then to put the planet in the plane of the ecliptic (where all other planets are), at the distance predicted by the Titius-Bode rule of 38 AU (see Box 2, "Titius-Bode rule"). Finally, he presented a solution to an inverse problem of determining the orbit of the trans-Uranian planet, by minimizing the residuals of the predicted and observed locations of Uranus. Le Verrier's solution was elegant and authoritative, in words of Airy (1846)[1]: "*It is impossible, I think, to read this letter without being struck with its clearness of explanation, with the writer's extraordinary command, not only of the physical theories of perturbation but also of the geometrical theories of the deduction of orbits from observations, and with his perception that his theory ought to explain all the phenomena of the planet's place*"

Le Verrier concluded his paper with a prediction of the location of the trans-Uranian planet at 01 January 1847 (325º of heliocentric longitude) and estimated an error of about 10º. This was a rather large error, but the paper delivering such a sensational claim was met with approval and applause. While everybody was impressed, nobody wanted to put it to test and look for the planet - or so it was thought.

**Controversy and theft**
While reading Le Verrier's June paper, Airy knew something nobody else did: Le Verrier's prediction was remarkably similar to a prediction of another young mathematician: John Couch Adams. The story, as it is usually told (e.g. Standage 2000), before the new evidence resurfaced in 1999 (Kollerstrom 2003), is that Adams started working on the Uranus motion soon after graduation in 1843 and by September 1845 had a solution for an orbit of a trans-Uranian planet, which he told to James Challis, the Plumian professor of astronomy in Cambridge and the Director of the Cambridge Observatory. Challis put Adams in contact to Airy and, as the story goes, Adams made 2 unsuccessful visits to Greenwich, each time missing Airy, but at lest leaving a note with a possible position of the planet.

---

[1] The quotation actually referrers to a letter from Le Verrier to Airy on 28.06. 1846, which answers Airy's question (in a letter from 26.06.) on the solution of the radius vector, but I believe it can be applied to general impression by Airy on Le Verrier's work.

How precise was the prediction of the position of a new planet on the note Adams left, and how it would fare in a comparison with the one of Le Verrier's June 1 paper, is difficult to demonstrate. The reasons for this are discussed extensively by Rawlins (1992), Sheehan et al. (2003; 2007) and Kollerstrom (2006a; 2006b; 2009)[2]. These authors point out that the note claimed to be Adams' prediction of September 1845 might actually be of a much later date. The date on the note supposedly left by Adams at Greenwich is vague (October 1845) and written in a different handwriting to the rest of the message; it is imprecise in explaining what kind of calculations had actually been done and it gives the *mean heliocentric longitude* as 325° 2' degrees, which, when converted to the true heliocentric longitude at the day of discovery, is 328° 41'. This value should be compared to the actual location of Neptune on that day, 326° 57'. Le Verrier's first prediction (June 1) was 324° 35', while in the final paper on 31 August, the one used by Galle and d'Arrest, he both improved the prediction to 325° 58' and declared that the planet should be recognisable as a disc (Le Verrier 1846b). In contrast, all other predictions by Adams were significantly worse than (supposedly) his first one (Rawlins 1992; Kollerstrom 2006).

Even though Adams' prediction turned out not to be as accurate as Le Verrier's, it was an amazing achievement. That was not lost on Airy (a master of celestial mechanics and a former Plumian Professor), but he was sceptical and wanted to see if Adams can also explain his 1838 discovery of the change in the radial motion of Uranus. Adams, however, did not answer Airy's inquiry in November 1845; this was the same question Airy asked Le Verrier in June 1846 and got an immediate answer that impressed him so much. Adams himself also never published anything of his (pre-discovery) calculations until November 1846 (Adams 1846b), even though he was a member of the RAS and had previously published a notable paper on the trajectory of a comet (Adams 1846a). Finally, the works cited above stress that the whole British claim of Adams' prediction, supposedly predating Le Verrier's work by some nine months, was actually put forward *after* the discovery of Neptune.

The post-discovery claim might be even taken as a full blown conspiracy theory, especially as all documents of the Royal Greenwich Observatory pertaining to the discovery of Neptune disappeared for more than 30 years (Kollerstrom 2003). When scholars started asking for some of the files (Chapmann 1988; Rawlins 1992) they were told that they were not in the RGO library, but gone missing.

In 1999 they resurfaced in Chile, among the possessions of recently deceased astronomer Olin Eggen (together with another large quantity of 17th century manuscripts and 60 rare books). It seems Eggen "borrowed" the Neptune files, as they are usually called, in order to write essays on Airy and Challis, probably while he was working at the RGO as an assistant to the Astronomer Royal in 1964. He never returned his loans to the library, moving them first to Australia then to Chile, straightforwardly denying of having these files as late as in 1996[3].

---

[2] See also: http://www.dioi.org/kn/neptune/index.htm.
[3] See: http://www.dioi.org/kn/neptune/takes.htm

**A secretive search and an international scandal**

After reading Le Verrier's June paper, and having received the explanation to the radial motion of Uranus in a letter directly from Le Verrer, Airy was so impressed that he thought the time had come to stir Challis into action. Airy devised a way to search for the planet centred on the location of Le Verrier's prediction. This is an interesting point: a director of the most prestigious observatory in the world was not actually starting a search for the planet himself, the largest prize in astronomy of the day, but outsourced the search, even offering a reliable assistant for help. As put forward by Chapman (1988) and Smith (1989), it is likely that Airy could not imagine interrupting RGO's very public duties, but was more than happy to set up the discovery of the age for Cambridge and its observatory.

Challis indeed started a rather secretive search on July 29. There is evidence that the search was kept secret from even his British fellow astronomers (see Rawlins 1992, for example). Unfortunately, even though the planet was observed three times, it was not recognised as such (see Box 3, "The failed attempt"). After the news of the discovery was circulated in Britain by Joseph Hind, Sir John Herschel was first to announce the (co-)prediction of Adams (Herschel 1846), while on October 17, Challis and Adams (Challis 1846b), using all of the available observations of the new planet, determined its distance and proposed a name for it: 'Oceanus' (see Box 4, "Naming of the planet").

The reaction in Paris can easily be guessed. Two weeks before, Le Verrier was the one person who "discovered the planet with the point of his pen"; not even Galle considered himself a co-discover, but just a person who found it. Suddenly and totally unexpected, there was another claim, with no actual proof, that supposedly predated Le Verrier's work, and assumed enough credit to take the honour of naming the new planet. Transporting the scene to a Jane Austen novel and one can easily imagine Paris Observatory director François Arago fuming and pacing in the shrubbery exclaiming: "Is it to be endured? But it must not, shall not be." This is what he did, but not, however, in "a prettyish kind of little wilderness" on one side of the lawn, but in the hall of the Académie des Sciences. His audience was the cream of the Parisian scientific establishment and journalists, and Arago proclaimed he would forever call the new planet "Le Verrier". The press was more than happy to take it from there and made an international scandal out of it.

Louis Pasteur is accredited with saying that science knows no nationality, only scientists do. The case of the discovery of Neptune adds another layer to it: scientific results have a tendency to be wrapped in the national (university, institute, or funding body) flag. Early historians have either struggled to understand or ignored Airy's writings (e.g. see Smart 1946a,b), supporting the British claim for co-prediction. A reanalysis of the historical events by Rawlins (1992) and the evidence coming from the files resurfaced in 1999, as presented by Kollerstrom (2006), showed clearly that Airy was building a case for British (and Cambridge) role in the discovery of Neptune. Once there was a rigorous prediction where to look (and a confirmation of a less rigorous but nevertheless indicative estimate), he pushed for the search, which unfortunately did not result in a foremost discovery. Airy's plan misfired and in the post-discovery national fervour it was Airy (together with Challis) who was blamed for the failure in an event that overshadowed the rest of his illustrious career.

**Two co-discoverers**
In a private letter to Le Verrier on 14 October 1846, Airy wrote: "*You are recognised beyond doubt as the real predictor of the planet's place*", and in his November 13 address to the RAS (Airy 1846), he compared Le Verrier's work as nothing as "*so bold... in astronomical prediction*" since Copernicus, concluding "*it is here that we see the philosopher*" (rather than just a mathematician). But Airy also called the discovery: "*the movement of the age; ... it has been urged by the feeling of the scientific world in general, and has been nearly perfected by the collateral, but independent labours, of various persons possessing talents or powers best suited to the different parts of the researches.*"

In both cases Airy is right. Le Verrier in three rigorous papers solved the problem of the motion of Uranus, and openly put his name behind a theory, for good or worse. It was he who urged the observers to test his prediction, an opportunity that most people rejected, or attempted in lukewarm fashion (e.g. at the Paris Observatory). The credit for the prediction has to go to him.

Airy is also correct in his assessment that this was the "movement of the age". This is especially true in modern science, where many people work on similar topics and simultaneous or nearly simultaneous solutions or discoveries happen often. The problem of Uranus was certainly one of the top problems in astronomy of the first half of 19th century. Le Verrier was told by Arago that he should have a look at it. Adams got inspired reading about the problem of Uranus in a report by Airy and about the perturbation theory in the 6th edition of Marry Somerville's "On Connexion of the Physical Sciences" (Chapman 2016). The uncertainty of Adams' predictions (spanning more than $20^o$), which had an unfortunate effect of misdirecting the secret search, cannot be used to simply dismiss Adams. He did work on the theory of Uranus, he might even have had a comparable solution, but he never went public with his prediction, had difficulty settling on the final position and, essentially, did not influence the discovery in the least.

On the other hand, there was a co-discover who certainly did play a major part, but whose credit was slow in coming. d'Arrest was present during the observations, it was his idea to look for the new charts, and he was checking the stars on the map. It is he who exclaimed: "*That star is not on the map!*" (Dreyer 1882). When Encke, as the director of Berlin Observatory, sent a letter to the Astronomische Nachrichten announcing and describing the discovery, he failed to mention d'Arrest at all. Almost nothing was known about his role until some 30 years later. In the mean time d'Arrest became a famous astronomer in his own right (see Box 5, "The discoverers). When d'Arrest was awarded the Gold Medal of the RAS in 1875, in the address delivered by none other than RAS President John Couch Adams there was no mention of d'Arrest's role in the discovery of Neptune; the Gold Medal was awarded for his work on nebulae (Adams 1875). The obituary published in the Monthly Notices also makes no connection between d'Arrest and Neptune.

Still, there were people who knew d'Arrest better. In a German obituary by J. E. L. Dreyer (Dreyer 1876) there is a sentence declaring d'Arrest participation in the discovery. Motivated perhaps by these oversights, Galle himself wrote two descriptions of the discovery (Galle, 1877, 1882) in which he gave credit to d'Arrest. Another influential revelation was the publication of Dreyer (1882), in which he described observing with d'Arrest's in 1874, when d'Arrest retold his memories of the night of the famous discovery.

Why was d'Arrest initially neglected? One should probably take into account the spirit of the age, when discovery announcements were short letters to the editor of a journal and the directors of observatories reported what their nameless assistants discovered. Galle, already an established astronomer, featured prominently in Encke's report (Encke 1846), but the mere student d'Arrest was not mentioned at all. Wolfgang Dick showed that Encke was later actually sorry not to include d'Arrest in the report and expressed his misgivings in a letter to Otto Struve, the director of the Pulkovo Observatory (Dick 1985, 1986).

d'Arrest's role in the discovery of Neptune is now securely known, but recognition came late. The naming of the rings of Neptune (Guinan et al. 1982) serves as a reminder how perceptions change; they were named after the principle participants in this scientific drama. In order of distance from the planet the main rings are called: Galle, Le Verrier and Adams; fainter features also carry the names of William Lassel (discoverer of Neptune's moon Triton) and Arago. It seems that even at this time, d'Arrest's role was not widely known or appreciated.

**A happy accident?**
The prediction of the position of Neptune by Le Verrier was an astonishing and inspirational application of a theory, demonstrating the power of science. It is a wonderful story made very human with the controversy of who-did-it-first, the naming scandal, the press war, the theft of crucial documents and the recent re-evaluation of the British contribution. Yet there is even more in this drama. Having two (unrecognized) pre-discovery and one (unrecognized) post-discovery observations by Challis, spanning some six weeks, Adams was able to calculate the new orbit of Neptune (Challis 1846a). In the new orbit, Neptune turned out to be much closer than predicted by Titius-Bode rule, at 30 AU, and closer than his and Le Verrier's solutions required. The data still did not allow for a more robust estimate of the eccentricity of the orbit; a larger time span was needed for this.

American astronomer Sears Cook Walker, working at the US Naval Observatory, read Le Verrier's publication of June 1846 and suggested to his superior officer that they should start a search for the planet. This was rejected because of the busy observatory schedule. When the news of the discovery steamed into the Boston harbour onboard *SS Caledonia* on 20 October 1846, the search for the planet was no longer necessary, but Walker recognised the importance of examining if there were, as in the case of Uranus, previous observations of Neptune (Table 1). Indeed, Walker discovered that J.J. Lalande's well known "Historie céleste française" contained an observation of a star that was consistent with the known orbit of Neptune, but was not in subsequent catalogues, and crucially, it was not visible anymore on the sky (Hubbell & Smith 1992). Further investigation showed the observations consisted of two sightings on May 8 and 10, 1795, remarked as doubtful, as it seemed that the "star" moved. This gave a base line of more than 50 years, a sufficient period for calculating the orbit of Neptune. Walker's result was to stun the astronomical world.

The main orbital parameters of Neptune are its distance, period and eccentricity. Walkers calculation confirmed Adams' estimate of 30 AU for the mean distance, derived the eccentricity of 0.0088 and the period of 166 years. Both values were radically different from the Le Verrier's (and Adams') prediction (see Table 2 for comparison of orbital elements). The orbit was much more circular, and as it was closer, the period was also shorter. This was further taken by Benjamin Peirce, a Perkins Professor of astronomy and mathematics at the Harvard University, who confirmed Walker's result and publicly proclaimed that "*the planet Neptune is not the planet to which geometrical analysis had directed the telescope; ...; and that its discovery by Galle must be regarded as a happy accident*". Furthermore, Peirce noticed that the orbital periods of Uranus and Neptune are close to 1:2 ratio, implying that the two planets could be in near resonant orbits. What made Peirce's statement world famous is that he disputed the Le Verrier's orbit with the calculated period of 217 years. This period put Uranus and Neptune close to the 2:5 resonance; this would be likely to have very peculiar effects on the orbit of Uranus, which Le Verrrier had not taken into account. Pierce's position was that Neptune was not responsible for the perturbations of Uranus. After Pierce calculated the mass of Neptune based on the observations of the orbit of its moon Triton, he changed his opinion and proclaimed that Neptune can account for the perturbation of the Uranus' orbit, including the earliest recorded observation of Uranus from 1690 by Flamsteed, which always had the largest error in both Le Verrier's and Adams' calculations (for a detail discussion see Hubbel & Smith 1992).

But was it a chance discovery or not? Le Verrier's prediction put Neptune's orbit much further from the Sun, but only on average. The orbit also had a significant eccentricity of about 0.1. Moreover, at the time of discovery the predicted planet was essentially closest to the location of the actual planet, at about 33 AU (Rawlins 1992; Kollerstorm 2006). As Danjon (1946) showed (Figure 2), both Le Verrier and Adams had to construct orbits such that they approached Neptune's orbit in order to minimise the terms of the discrepant Uranus's motion. Their calculations, while globally incorrect, did approach the actual position of Neptune on the sky.

**Modern insights**

A modern approach to the solution of the perturbations of Uranus was discussed in details by Lai et al. (1990), providing an insightful analysis of Neptune's influence on the orbit of Uranus. They solved both the forward and the inverse problem, respectively predicting the perturbations of Uranus given the modern orbital elements of Uranus and Neptune and determining the orbital elements of Neptune using the residuals between observed and predicted positions for Uranus. Lai et al. showed that the residuals of the Uranus motions depend on two dominant terms (Box 6, "Explaining Uranus' motion"): the force Neptune exerts on Uranus which is dependant on Neptune's mass and radius, $M_N/R^2_N$ (the *inhomogeneous* solution); and the difference between two Keplerian orbits, expressed as the orbit of Uranus with perturbed semi-major axis and eccentricity (the *homogenous* solution).

Neptune has a large pull and, if other contributions are removed, it would account for almost 550 arcseconds in the deviation of the Uranus position. At the time of the discovery the observed deviations were of the order of 50 - 100 arcseconds, about a factor of 5-10 less (Figure 1). This arises because the other term, that describing the shift in eccentricity of Uranus, has also an amplitude that would produce about 500 arcseconds of deviation if considered alone. Here is the crucial insight first indicated by Pierce: Neptune and Uranus are in near 1:2 resonance (less than 2% deviation), so the orbital periods introduce an important beat effect. As demonstrated by Lai et al., the phases of the two dominant terms are such that they nearly cancelled each other out in the early 1800s. Today or at the time of Galileo, the perturbation are constructive and result in much larger deviations.

The discovery of Neptune was not just lucky: as the predictions were solid. The inverse problem that Le Verrier and Adams attempted to solve has seven unknown elements: Neptune's orbital period, time of conjunction, Neptune's mass, and four constants of the homogeneous solution describing the true (perturbation free) orbit of Uranus. As Lai et al. show, a perturbed orbit of Uranus can also be described as a unperturbed orbit with a modified eccentricity. Thus, understanding the perturbation Neptune exerts on Uranus by its mass and radius is made difficult by the degeneracy between the unknown true orbit of Uranus (if Uranus were alone in the Solar system) and a perturbed one of slightly different orbital parameters.

**One year on Neptune**
The discovery of Neptune took place 170 years ago, just a little more than it takes Neptune to make one revolution around the Sun. This one Neptune year has brought major changes in both human society and science. The distribution of information is now essentially instantaneous, something that Otto Struve would have valued tremendously. He also received a letter from Le Verrier, sent on the same day as the one to Galle, but it arrived 6 days later to Pulkovo Observatory near St. Petersburg, by which time the discovery had already been announced (Dick 1986). Nowadays it is unthinkable to submit an observational proposal not supported by some kind of theoretical predictions, while Le Verrier struggled to persuade observes to look. The distribution of orbits of trans-Neptunian bodies shows tantalising evidence for a ninth planet (Trujillo &

Sheppard 2014; Batygin & Brown 2016), and the Solar System now looks very different from that known to Le Verrier, Galle, d'Arrest and Adams.

Some things, however, do not change. A discovery requires deep knowledge, bold thinking and some luck. The luck was absent in Cambridge, but Le Verrier's dauntless audacity, as well as Galle's and d'Arrest's willingness to take the challenge, should be celebrated. The discovery of Neptune is a quintessential story about progress in our understanding of the universe, and also about how science works in a socio-political context. It is a story worth remembering and a good way to engage the general public in a dialogue about science.

*Acknowledgements. I would like to thank librarians Regina von Berlepsch, Marcel Thies (AIP) and Sian Prosser (RAS) for providing many useful articles, and Matthias Steinmetz for stimulating my curiosity to develop this work. I am grateful to Robert Matthews for spotting a mistake in Table 1 regarding the date of Galileo's observations.*

**BOX 1. The chart**

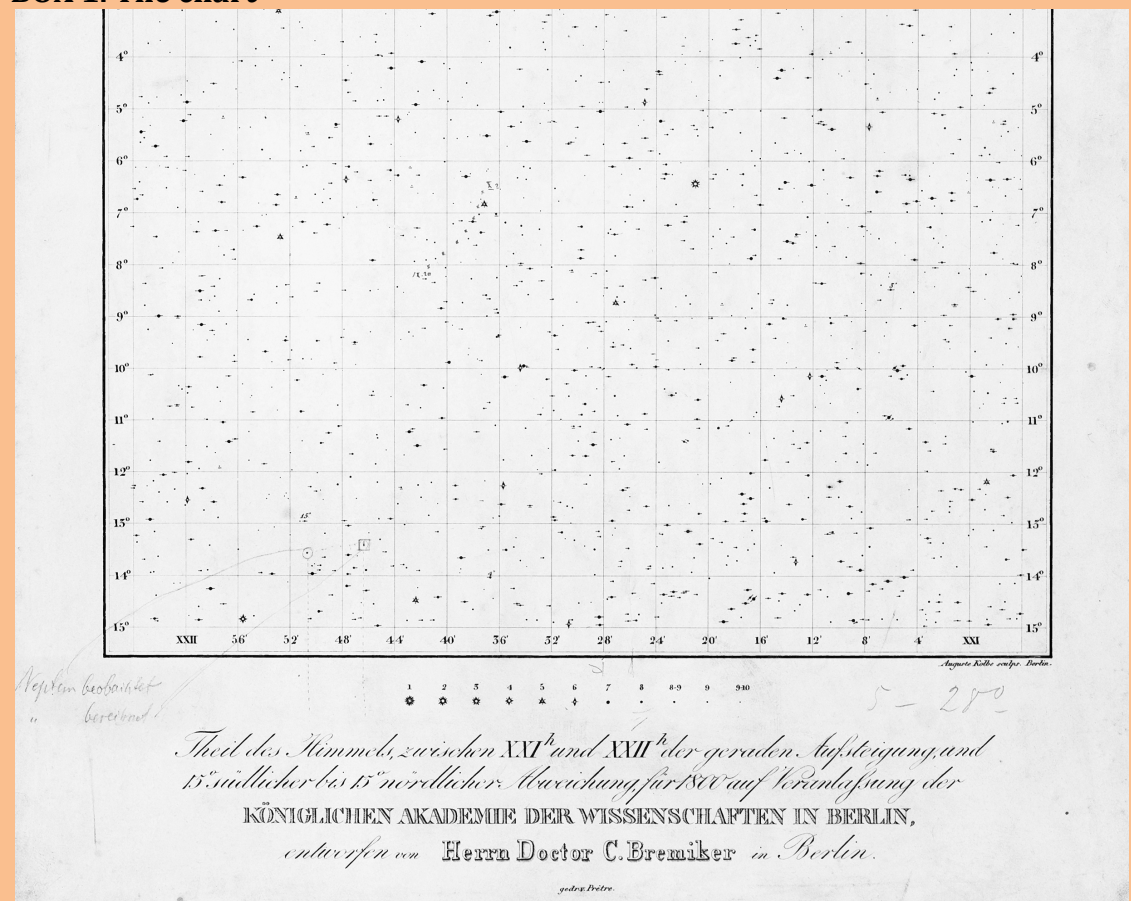

A partial scan of the chart Hora XXI used by Galle and d'Arrest in their search for Neptune. It was produced by Carl Bremiker at the Berlin Observatory for the Royal Academy of Sciences in Berlin. Bremiker produced 4 other charts (Hora VI, IX, XIII and XVIII), more than any other astronomer in that series. In the lower left corner there is a square and a circle, showing the predicted ("Neptun berechnet") and observed ("Neptun beobachtet") positions of Neptune, respectively. (Library of the Leibniz-Institut für Astrophysik Potsdam.)

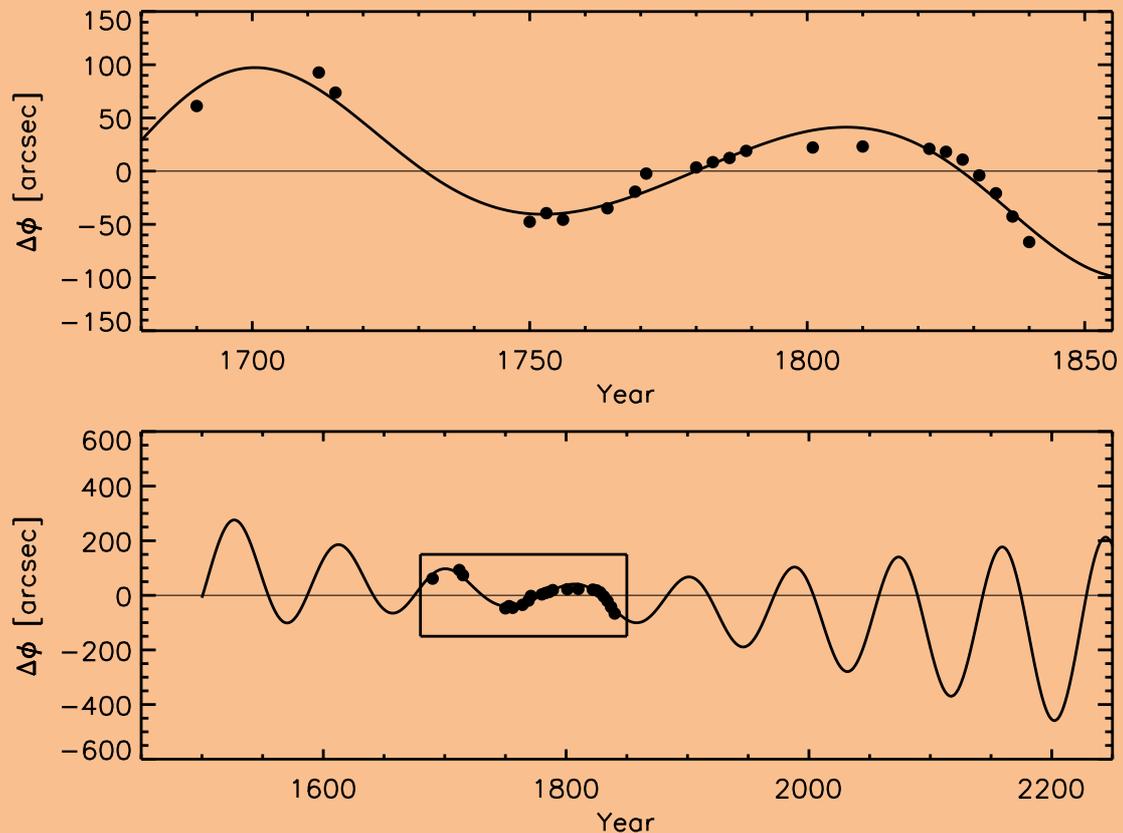

**Figure 1. Uranus out of control.** Discrepancy in predicted and observed heliocentric longitude of Uranus (points) and the fit to the model. The upper panel shows the measurements from the period used for the prediction by Le Verrier and Adams, while the lower panel shows the predicted residual on a longer timescale assuming the same model that fits the historic data. The boxed region in the lower plot corresponds to the upper plot.

**BOX 2. Titius-Bode rule**
Titius-Bode rule (or law) describes a fact that the distance of a planet from the Sun follows a sequence, expressed as a formula *a =0.4+0.3×2$^n$*, where *a* is the distance in AU, and *n* increases by 1 for each planet, starting with -∞ for Mercury and 0 for Venus. It was introduced by Johan Daniel Titius in the preface to his translation of "*Contemplations de la Nature*" by Charles Boinnet, and Johan Elert Bode, the predecessor of Encke as the director of the Berlin Observatory, who was almost evangelical in advertising it. When it was discovered that Uranus also fits to the rule (with n=6), and that multiple minor planets circle at the value for n=3 (between Mars and Jupiter), Titius-Bode rule moved from being a curious fact to a main stream astronomy tool. Crucially, it was used by Le Verrier and Adams as a starting estimate for the distance of the perturbing planet, breaking a degeneracy between mass and distance (~M/R$^2$). The discovery of Neptune eventually showed the non-universality of this rule.

> **BOX 3: The failed attempt**
>
> The only observatory to take Le Verrier's initial prediction seriously and mount a systematic search to find the new planet was the Cambridge Observatory under directorship of James Challis (1803 -1882). The *spiritus movens* of the search was, however, Geroge Biddel Airy (1801 - 1892), Astronomer Royal, who got convinced that there indeed could be a planet having seen both Le Verrier's and Adams' predictions. He pushed Challis to do the search, proposed a method and the area of the search around Le Verrier's prediction, and sent help in form of an assistant observer from the Royal Greenwich Observatory. The role of John Couch Adams (1819 - 1892) for the search was crucial. Not only did he first predict the existence of a perturbing planet in 1845, but during the search he provided several other possible locations of the planet. Unfortunately, they were mostly inconsistent with each other, swinging some 20 degrees and sending the search in wrong directions (Rawlins 1992). Challis' search is also infamous for having observed Neptune three times, but not recognising it as a planet. When he was checking the validity of the search method, comparing observations between the nights of 4 and 12 August, he stopped at star number 39, being satisfied that the method was working. If he had continued but a bit longer, Challis would have no doubt noticed that the entry number 49 of the 12th changed its position since the 4th. The Cambridge search was essentially a total failure, and in the post-discovery assessment two legends were born.
>
> The first one relates to the fact that Challis lacked Bremiker's *Hora XXI* chart that Galle and d'Arrest used at the Berlin Observatory. This is, of course, true, but it is remarkable that Challis had *Hora XXII* chart, adjacent and partially overlapping the map in Berlin. As Kollerstorm (2006a) noticed, during August Neptune was on the map Challis had. The other legend is related to the fact that Challis told an assistant to note next to an entry in the logbook: "*The last one seemed to have a disc*". This was indeed, as Challis later found out, Neptune and not a star. The first part of the note "*The last one*" was crossed over, probably post-discovery, because Challis never stopped his telescope to verify the claim, even though he was by then aware that Le Verrier advocated looking for a disc.
>
> The Neptune affair had profound implications on the careers of the main participants. It completely overshadowed Airy's and Challis' work, but created from Adams a star. The re-assessment of the British part in the discovery of Neptune, however, paints quite a different picture, especially of Airy and his crucial role in both establishing the search for the planet and building the British claim for co-discovery (Kollerstrom 2006a).

**Table 1. Pre-discovery sightings of Neptune***

| Date of observation | Observer | Discoverer |
|---|---|---|
| 28 Dec 1608 & 27 Jan 1609 | G. Galilei | Kowal and Drake (1980) |
| 08 and 10 May 1795 | M. Lalande | S.C. Walker, A.C. Petersen, F. Mauvais (1847) |
| 25 Oct 1845 | J. Lamont | J. Hind (1850) |
| 04 and 12 Aug 1846 | J. Challis | J. Challis (1846) |
| 7 and 11 Sept 1846 | J. Lamont | J. Hind (1850) |

*Data taken from Rawlins (1992)

**Table 2. Orbital elements of Neptune***

| orbit | Le Verrier | Adams | Walker | Neptune |
|---|---|---|---|---|
| semi-major axis [AU] | 36.15 | 37.25 | 30.25 | 30.11 |
| discovery distance [AU] | 33 | 32 | - | - |
| eccentricity | 0.10761 | 0.12062 | 0.00884 | 0.009456 |
| orbital period [yr] | 217.4 | 227.3 | 166.4 | 164.8 |
| mass [$M_{SUN}$] | 0.00011 | 0.00015 | 0.000067 | 0.0000515 |

*Comparison of pre- and post-discovery orbital elements. For Le Verrier, Adams and Walker elements data are taken from Grosser (1962), and the discovery distances from http://www.dioi.org/kn/neptune/witihin.htm.

**Table 3.** Comparison of the fits of the Lai et al. (1980) and in this work.

|  | γ | $\beta_1$ | $\beta_2$ | $\beta_3$ | $\beta_4$ |
|---|---|---|---|---|---|
| Lai et al. | 890" | -18.1" | -45.4" | 841" | 76.8" |
| This work | 550.36" | -8.48" | -13.59" | 504.63" | 33.25" |

**Figure 2. Orbits of Neptune.** A schematic description of the orbits of Uranus and Neptune and the predictions by Le Verrier and Adams. Note different eccentricities in the predicted orbits, their mutual similarities and the approach to the true orbit of Neptune around the time of discovery. (From Danjon 1946).

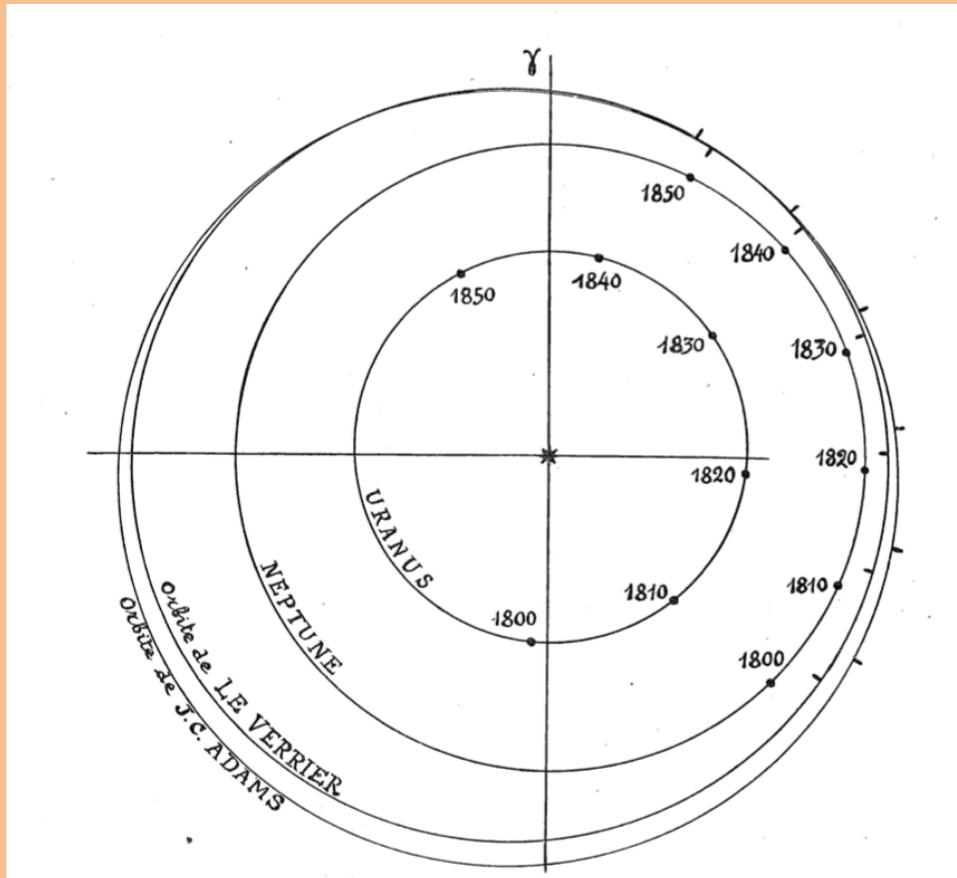

**BOX 4 Naming the planet *

Finding the name for Neptune was extremely quick, but agreeing on using it, as almost anything in the story of its discovery, was a scandalous affair (e.g. Kollerstrom 2009). In the letter announcing the discovery of the predicted planet, Galle, assuming that he could put forward a name as a person who found it, suggested the name "Janus", the Roman god of beginnings and passage. Le Verrier, who in his announcement of the discovery in the French newspapers already suggested "Neptune", immediately rejected Galle's suggestion, saying that the name was chosen by the Bureau des Longitudes, the latter actually not being true. In the various announcements that followed across the continent, like the one in Britain by Joseph Hind in The Times, the planet was often called "Le Verrier's planet". The real storm was started by a written suggestion by Challis and Adams in the Athenaeum the planet should be called "Oceanus". J. Hind considered this an unfortunate choice writing "*it is no more likely to succeed with the French (who have the only right to name it) than if it had been dubbed 'Wellington'*". Indeed, the choice upset the French, and Arago proclaimed that he will always call it "Le Verrier" (and Uranus "Herschel"). Le Verrier found himself in an awkward situation of suggesting one name, but using the one Arago proclaimed. Sir John Herschel proposed a few other mythological names: "Demogorgon", "Minerva" and "Hyperion". "Neptune", however, stuck in the minds of most continental astronomers and Airy adopted it in early 1847. He might have been following the reasoning of the president of the RAS, Captain W.H. Smyth, who forgetting that a German *did* discover both Uranus and Neptune wrote: "*I don't quite like this proposed change in the nomenclature of the Planets, for mythology is neutral ground. Herschel is a good name enough. Le Verrier somehow or other suggests the idea of a Fabriquant & is therefore not so good. But just think how awkward it would be if the next planet should be discovered by a German: by a Bugge, a Funk, or your hirsuite friend Boguslawski!*". The choice of the mythological name had one important consequence, it finally convinced the editors of the *Nautical Almanac* to adopt the name Uranus, instead of "The Georgian".

*Quotes are taken from http://www.dioi.org/kn/neptune/corr.htm

**BOX 5: The discoverers**

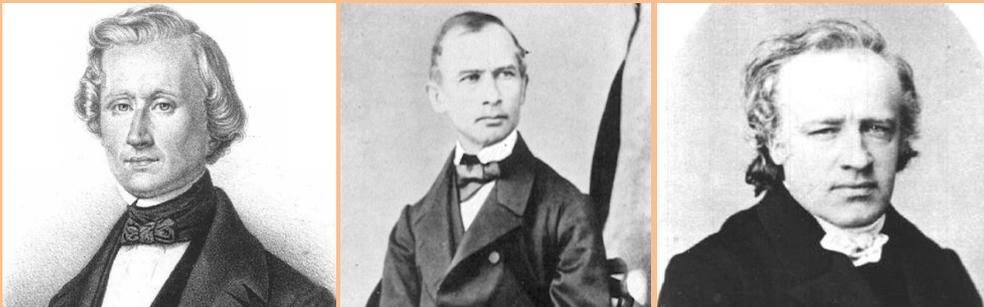

Le Verrier found Neptune "with the point of his pen", but it was Galle and d'Arrest who identified it in a telescope.

**Jean Joseph Urbain Le Verrier** (1811 - 1877) was born in Saint-Lô, Normandy, and started as a chemist, but got a position of assistant professor of astronomy at the École Polytechnique, switching his interests to celestial mechanics. He worked on the stability of the Solar System and orbit of Mercury before Arago saw in him the right person to tackle the motion of Uranus. After the discovery of Neptune, he returned to the Mercury problem, and, not willing to give up on Newtonian gravity, predicted the existence of another planet close to the Sun, or at least a belt of smaller bodies, within Mercury's orbit. In 1859 an amateur astronomer E. Lescarbault announced a sighting of such a planet; it was quickly called Vulcan by the press, but was never seen again. For Mercury, it was the law of gravity that needed adjustments, as Albert Einstein showed in 1915. Le Verrier served as a director of the Paris Observatory, until he was fired for rash treatment of his assistants, but was reinstated after the following director died. Le Verrier died on 31. anniversary of the discovery of Neptune at the Paris Observatory.

**Johann Gottfried Galle** (1812 - 1910) was born in Pabsthaus, about 100km south of Berlin. He went to a gymnasium in Wittenberg and attended the university in Berlin. He was a gymnasium teacher, before getting hired as the first employee (assistant to the director) of the new Berlin Observatory. He discovered the C ring of Saturn, but became famous with the discovery of three comets in consecutive months in 1839-1840 (Wattenberg 1963). For a while he was considered as a suitable successor of Friedrich Bessel at the Köningsberg Observatory, but eventually moved to Breslau (Wroclaw) as the director of the observatory and professor of mathematics. In 1872 he proposed a new method of measuring the solar parallax using asteroids and organised a world wide observations of Flora's transit. Galle died in Potsdam, a month past his 98 birthday.

**Heinrich Louis d'Arrest** (1822 - 1875) was born in Berlin where he studied mathematics and eventually joined the Berlin Observatory, sleeping in an attic room. In 1848 he moved to the Leipzig Observatory, where he later became an adjunct professor at the university, a title he received in return for not taking a post in Washington. In 1852 d'Arrest moved to Copenhagen as the professor of Astronomy and head of the observatory. d'Arrest discovered several comets and an asteroid (76) Freia. In Copenhagen he started working on nebulae including the external galaxies (especially in the Coma Cluster). d'Arrest died in Copenhagen.

**Box 6: Explaining Uranus' motion**. When influences of all known planets were taken out, Uranus showed a notable discrepancy between the predicted and observed position, $\Delta\phi$. Figure 1 shows the heliocentric longitude data points similar to those used by Le Verrier and Adams in their calculations (as presented in Lai et al. 1990). The line is the solution of the forward model, which takes into account the known orbital elements of Neptune and Uranus, and is given by the following equation (eq. 19 of Lai et al.):

$$\Delta\phi = -\gamma \sin\left[2(\Omega_U - \Omega_N)\tau\right] + \beta_1(\Omega_U - \Omega_N)\tau + \beta_2 + \beta_3 \sin(\Omega_U \tau) + \beta_4 \cos(\Omega_U \tau)$$

where $\Omega_U$ and $\Omega_N$ are angular velocities of Uranus and Neptune, respectively, and $\tau = t - t_0$ is the time with respect to the year of conjunction ($t_0$ = 1822). The first term describes the perturbation on Uranus due to Neptune's mass and radius, while the last four terms describe the difference between two nearby Keplerian orbits of Uranus and Neptune. Figure 1 also shows a new fit of the equation to the data, with somewhat different results to Lai et al. (see Table 3), but the trends are the same.